\documentclass[a4paper]{jpconf}
\usepackage{amsmath}
\usepackage{amssymb}
\usepackage{subcaption}
\usepackage{hyperref}
\usepackage{blkarray, bigstrut}

\usepackage{graphicx}
\graphicspath{{./graphics/}}

\DeclareMathOperator*{\E}{\mathbb{E}}

\begin{document}

\title{Machine Learning on sWeighted data}
\author{M Borisyak$^{1}$ N Kazeev$^{1,2}$}
\address{$^1$ Laboratory of Methods for Big Data Analysis, National Research University Higher School of Economics, 3 Kochnovsky Proezd, Moscow 125319, Russia}
\address{$^2$ Sapienza University of Rome, Piazzale Aldo Moro, 5, 00185 Roma RM, Italy}

\ead{nikita.kazeev@cern.ch}

\begin{abstract}
Data analysis in high energy physics has to deal with data samples produced from different sources. One of the most widely used ways to unfold their contributions is the sPlot technique. It uses the results of a maximum likelihood fit to assign weights to events. Some weights produced by sPlot are by design negative. Negative weights make it difficult to apply machine learning methods. The loss function becomes unbounded. This leads to divergent neural network training. In this paper we propose a mathematically rigorous way to transform the weights obtained by sPlot into class probabilities conditioned on observables, thus enabling to apply any machine learning algorithm out-of-the-box.
\end{abstract}

\section{Introduction}
\begin{figure}[ht]
    \begin{subfigure}{.49\textwidth}
        \centering
        \includegraphics[width=\textwidth]{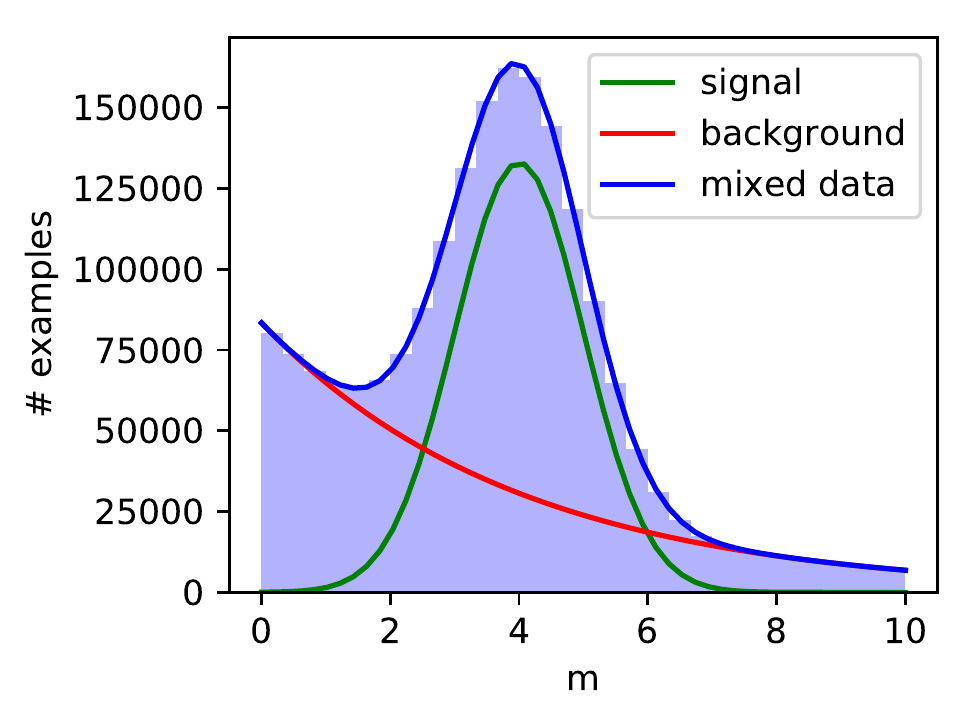}
    \end{subfigure}
    \begin{subfigure}{.49\textwidth}
        \centering
        \includegraphics[width=\textwidth]{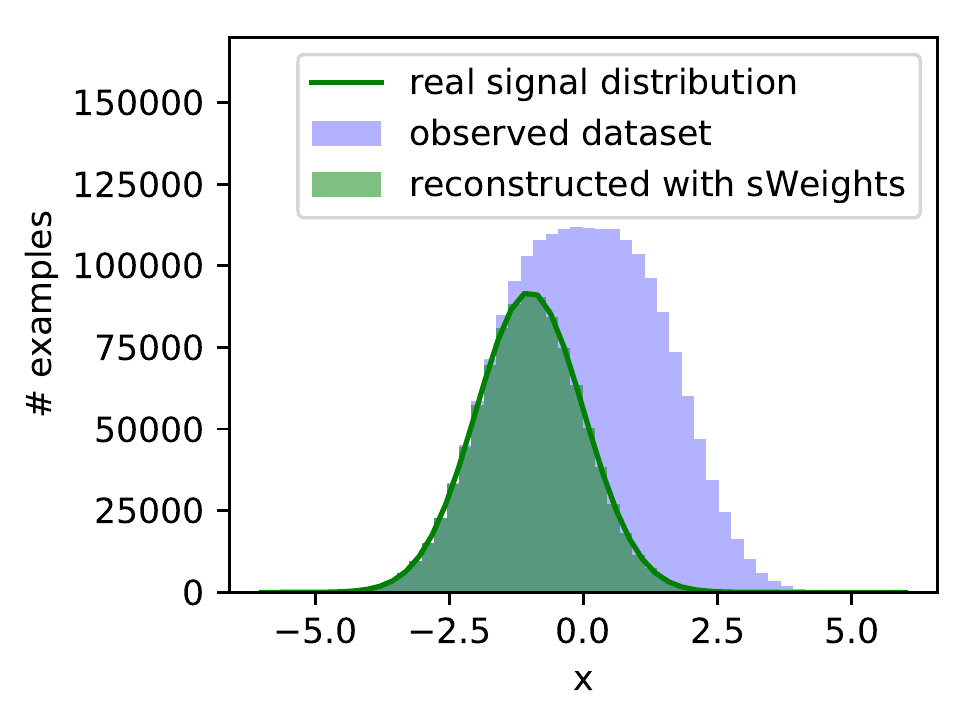}
    \end{subfigure}
    \caption{Demonstration of the sPlot technique. The known distributions of a discriminative variable $m$ (left) is used to weight the distribution of a target variable $x$ for mixture of signal and background events (right). Adapted from \cite{our-glorious-arxiv-paper}.}
    \label{fig:sWeights}
\end{figure}

Data obtained by high energy physics experiments is often a mixture of signal and background events. Application of conventional classification methods to such data is a challenging task as obtaining precise label information is usually impossible. Instead, one can use a probabilistic approach by employing so-called discriminative variables (usually, the invariant mass), whose distributions for signal and background events are known in advance, or can be estimated by a maximum-likelihood fit. The sPlot technique \cite{splot} allows us to recover the distribution of target variables for signal events, given that target variables are statistically independent of discriminative ones. sPlot achieves that by assigning weights (sWeights) to all events, e.g. histogram of the target variable is obtained as the sum of these weights within each bin:
\begin{equation}
    P(x \in Q \mid S) \approx \frac{1}{\sum_i w_i}\sum_i w_i \mathbb{I}[x_i \in Q],
\end{equation}
where $x_i$ are samples of the target variables, $w_i$ are weights assigned by sPlot, $Q$ denotes an arbitrary set and $S$ a signal event. Figure \ref{fig:sWeights} provides an illustration of sPlot technique. Derivation and detailed analysis of the sPlot are given in \cite{splot}.

A notable property of the sPlot technique is that these weights can be negative. Typically, This does not create problems for low-dimensional analysis with simple models. Nonetheless, one can encounter difficulties fitting large-capacity models, e.g. deep neural networks.
The prior work \cite{our-glorious-arxiv-paper} describes such difficulties for the problem of signal/background separation and introduces two approaches for dealing with them. In this work, we expand the previous results to the problem of classification between two signal sources, samples of each are contaminated by different sources of background events.

The rest of the paper is structured as follows. In the next section, we briefly outline prior results from \cite{our-glorious-arxiv-paper}. Section \ref{sec:we} introduces methods for signal-signal classification. Experimental results are presented in section \ref{sec:experiments}.

\section{Background}
\label{sec:background}

Consider the problem of training a classifier $f(x)$ to separate background (denoted as $B$) from signal (denoted as $S$) events given unlabeled sample $\{(x_i, m_i)\}^N_{i = 1}$, where $x$ denote target variables, $m$ --- discriminative variables. We assume that $x$ and $m$ are statistically independent:
\begin{eqnarray}
    p(x, m \mid S) &=& p(x \mid S) p(m \mid S);\\
    p(x, m \mid B) &=& p(x \mid B) p(m \mid B);
\end{eqnarray}
and $p(m \mid S)$ and $p(m \mid B)$ are known in advance.

One can use a cross-entropy loss to train a classifier:
\begin{multline}
    \mathcal{L}(f) =
    \E_{x, m \sim p(x, m)} \mathrm{w}(m) \log f(x) + (1 - \mathrm{w}(m)) \log (1 - f(x)) = \\
        \E_{x \sim p(x)} P(S \mid x) \log f(x) + P(B \mid x) \log (1 - f(x)), \label{eq:loss}
\end{multline}
where $p(x, m)$, $p(x)$ denote probability density functions and $w(m)$ the weights assigned by sPlot. The last equality is due to the property of sPlot:
\begin{equation}
    \E_{x, m \sim p(x, m)} \mathrm{w}(m) = P(S~\mid~x) \label{eq:swprop}.
\end{equation}
Proof of this statement can be found in \cite{our-glorious-arxiv-paper}.

Loss \eqref{eq:loss} can be approximated with a finite sample as follows:
\begin{equation}
    L_1(f) = \sum_i \mathrm{w}_i \log f(x_i) + (1 - \mathrm{w}_i) \log (1 - f(x_i)). \label{eq:lossfinite}
\end{equation}

This approach is quite popular in the high energy physics community \cite{our-glorious-arxiv-paper}. One might notice, however, that some of the summands in \eqref{eq:lossfinite} might not have a lower bound in case of negative $\mathrm{w}_i$. Therefore, if classifier $f(x)$ can isolate a point $x_i$ with a negative corresponding weight $\mathrm{w}_i$ from the rest of the sample, loss \eqref{eq:lossfinite} can be made arbitrarily low, which leads to quick overfitting. This is often possible for complex models, e.g. for large-capacity neural networks.

Two options are introduced in \cite{our-glorious-arxiv-paper} for avoiding this problem. The first approach, instead of cross-entropy loss, suggests using mean square regression directly on weights, taking advantage of the property \eqref{eq:swprop}:
\begin{equation}
    L_2(f) = \sum_i (f(x_i) - \mathrm{w}_i)^2.
    \label{eq:constrainedMSE}
\end{equation}
and restricting values of $f(x)$ to $[0, 1]$, e.g. by representing it as $\sigma(g(x))$ where $\sigma(\cdot)$ is the sigmoid function, $g(x)$ is an unbounded regression model.

Another approach is to use the maximum likelihood principle and avoid sPlot altogether and use the known probabilities $P(S \mid m)$ directly:
\begin{equation}
    L_3(f) = -\sum_i \log \left[ f(x_i) p(m_i \mid S) + (1 - f(x_i)) p(m_i \mid B) \right].
    \label{eq:exactloss}
\end{equation}

A detailed analysis of these loss functions and formal proofs can be found in \cite{our-glorious-arxiv-paper}.

\section{Our approaches}
\label{sec:we}

In this work we extend approaches discussed above to the case of classification between two classes ($C_1$ and $C_2$), both of them represented as mixtures of signal (denoted as $S_1$ and $S_2$) and background (denoted as $B_1$ and $B_2$) events: $\{(x_i, m_i, y_i)\}^N_{i = 1}$, where $x_i$ are samples of target variables, $m_i$ samples of discriminative variables, and $y_i$ is  binary class indicator.

Consider cross-entropy loss between $S_1$ and $S_2$:
\begin{equation}
    \mathcal{L}_4(f) = P(S_1 \mid S_1 \cup S_2) \E_{x \sim S_1} \log f(x) + P(S_2 \mid S_1 \cup S_2) \E_{x \sim S_2} \log(1 - f(x)) \label{eq:losss}
\end{equation}
where $\E_{x \sim S} h(x)$ denotes the conditional expectation $E_{x}\left[ h(x) \mid S \right]$, $P(S \mid S_1 \cup S_2)$ --- the probability of $S$ given $S_1$ or $S_2$.

\begin{multline}
    \mathcal{L}_4(f) =
        \alpha \E_{x \sim S_1} \log f(x) +
        \beta \E_{x \sim S_2} \log(1 - f(x)) =\\
    \alpha \E_{x \sim C_1} \frac{P(x \mid S_1)}{P(x \mid C_1)} \log f(x) +
    \beta \E_{x \sim C_2}\frac{P(x \mid S_1)}{P(x \mid C_1)} \log(1 - f(x)) =\\
        \alpha \E_{x \sim C_1} \frac{P(S_1 \mid x, C_1) P(x \mid C_1)}{P(S_1 \mid C_1) P(x \mid C_1)} \log f(x) +
        \beta \E_{x \sim C_2}\frac{P(S_2 \mid x, C_2) P(x \mid C_2)}{P(S_2 \mid C_2) P(x \mid C_2)} \log(1 - f(x)) = \\
    \frac{\alpha}{P(S_1 \mid C_1)} \E_{x \sim C_1} P(S_1 \mid x, C_1) \log f(x) +
    \frac{\beta}{P(S_2 \mid C_2)} \E_{x \sim C_2}P(S_2 \mid x, C_2) \log(1 - f(x)); \label{eq:loss4}
\end{multline}
where: $\alpha = P(S_1 \mid S_1 \cup S_2)$, $\beta =  P(S_2 \mid S_1 \cup S_2)$. Therefore, loss \eqref{eq:loss4} can be approximated by:
\begin{equation}
    L_4(f) = \sum_i \frac{y_i p_1(x_i)}{c_1} \log f(x_i) + \frac{(1 - y_i) p_2(x_i)}{c_2} \log f(x_i); \label{eq:losslik}
\end{equation}
where $c_1$ and $c_2$ are estimates of $P(S_1 \mid C_1)$ and $P(S_2 \mid C_2)$ correspondingly, $p_{1, i}$ and $p_{2, i}$ are estimates of $P(S_1 \mid x, C_1)$ and $P(S_2 \mid x, C_2)$ obtained by e.g. exact likelihood approach from \cite{our-glorious-arxiv-paper}.

Alternatively, with the help of sPlot weights, loss \eqref{eq:losss} can be expressed through binary indicator $y$:
\begin{equation}
    \mathcal{L}_4(f) = \E_{x, y, m} y\cdot\mathrm{w}_1(m) \log f(x) + (1 - y)\cdot\mathrm{w}_2(m) \log(1 - f(x)),
\end{equation}
where $\mathrm{w}_1(m)$ and $\mathrm{w}_2(m)$ denote weights assigned by sPlot for classes $C_1$ and $C_2$ correspondingly.

Here we also encounter potentially negative weights and all the problems associated with them as discussed in section \ref{sec:background}. Applying weights averaging as in \cite{our-glorious-arxiv-paper}, one can avoid negative weights:
\begin{multline}
    \mathcal{L}_4(f) = \E_{x, y} \left(\E_{m} \left[\mathrm{w}_1(m) \mid x\right]\right) y \log f(x) + \left(\E_{m} \left[\mathrm{w}_1(m) \mid x\right]\right) (1 - y) \log(1 - f(x)) \approx \\
    \sum^N_{i = 1} y_i p_1(x_i) \log(f(x_i)) + (1 - y_i) p_2(x_i) \log(1 - f(x_i)),
\end{multline}
where $p_1(x)$, $p_2(x)$ can be estimated by a regression trained on the corresponding weights with the Сonstrained MSE loss (\ref{eq:constrainedMSE}).

\section{Experimental evaluation}
\label{sec:experiments}

We evaluate the proposed methods on data collected from the LHCb Muon subsystem \cite{archilli2013performance} (MuID). To our knowledge, this is the only open dataset \cite{IDAO} that has sWeights. The dataset contains different particle species obtained from different calibration decays \cite{lupton2016calibration} and is labeled. Each species has its own background that is subtracted with a separate application of the sPlot method. For simplicity, we use only pion and muon species and ignore the kinematic weights. We use 60 scalar features. We split the dataset into train and test parts containing $2\cdot 10^6$ and $6\cdot 10^5$ examples respectively. For each train dataset size, the classifier was fitted 10 times to estimate the standard deviation of the scores.

\begin{figure}[h]
    \centering
    \includegraphics[width=0.5\textwidth]{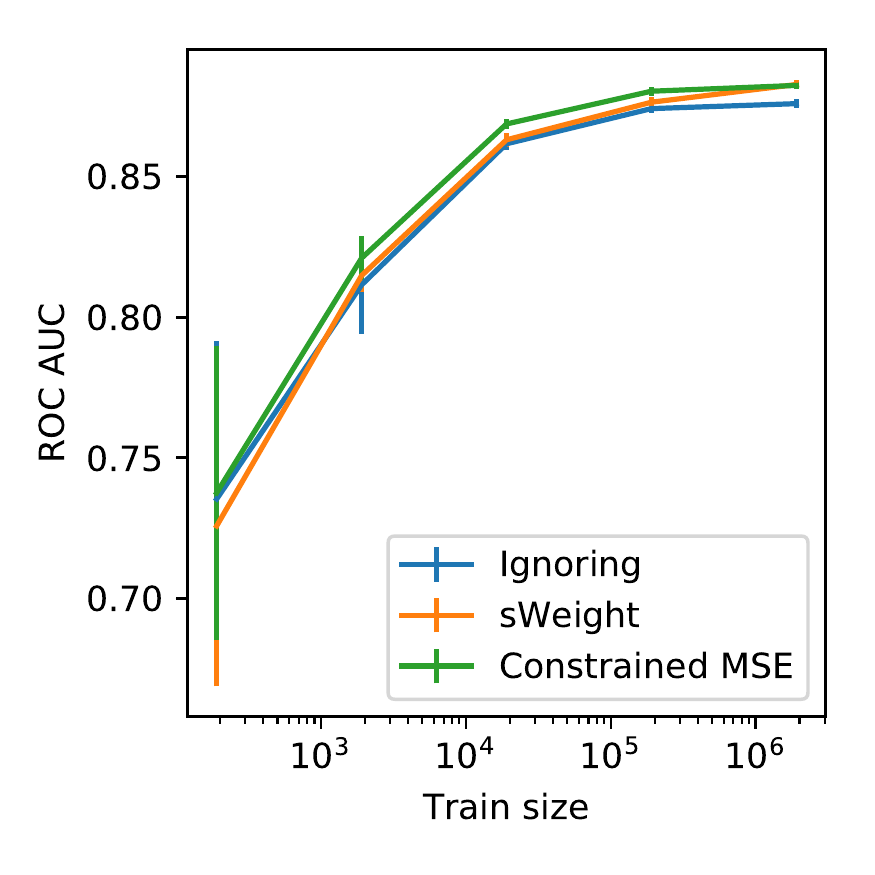}
    \caption{Experimental evaluation of CatBoost with the presented loss on the MuID dataset. sWeights -- using sWeights directly as weights for logloss; Constrained MSE -- our regression (\ref{eq:constrainedMSE}) used to transform the sWeights, the result used as weights for classification; Ignoring weight -- training ignoring the sWeights.}
    \label{fig:MuID-cb-raw}
\end{figure}

The results are on the figure \ref{fig:MuID-cb-raw}. The models used are described in \ref{sec:model_parameters}. Direct application of sWeights statistically significantly loses to our approach for some training sizes, performing about the same for the rest. The gap decreases with the training set size increase. However, ignoring the sWeights during training also yields good results, so it seems that the impact of sWeights on the problem is limited. This suggests, that the distribution of pionic background has good separation from the distribution of muon signal and vice versa. In this case, a classifier that is optimal for separating the muon signal/background mixture from the pion signal/background mixture is also almost optimal for the separation of muon signal from the pion signal. This limits the utility of the dataset with the respect to evaluating the proposed methods.

To address this problem, we do an additional test with synthetically constructed sWeights, that are tailored to impact muon/pion classification. There are many equally attractive ways to introduce such sWeights. We go with the most obvious one: mark a sample of particles that are muons with a high level of confidence as pion background. First, we train a CatBoost classifier to separate signal and background and a classifier to separate muons and pions as described in section \ref{sec:background}. We select 30\% of examples with muon label with the highest scores and marked them as pion background. Next, we use the signal probabilities obtained from the first step and assign definite signal and background labels by cutting on its output. We generate ``pseudomass" for both muon and pion background, each from a different distribution, which we use to compute the new sWeights.

\begin{figure}[h]
    \begin{subfigure}{0.5\textwidth}
        \includegraphics[width=\textwidth]{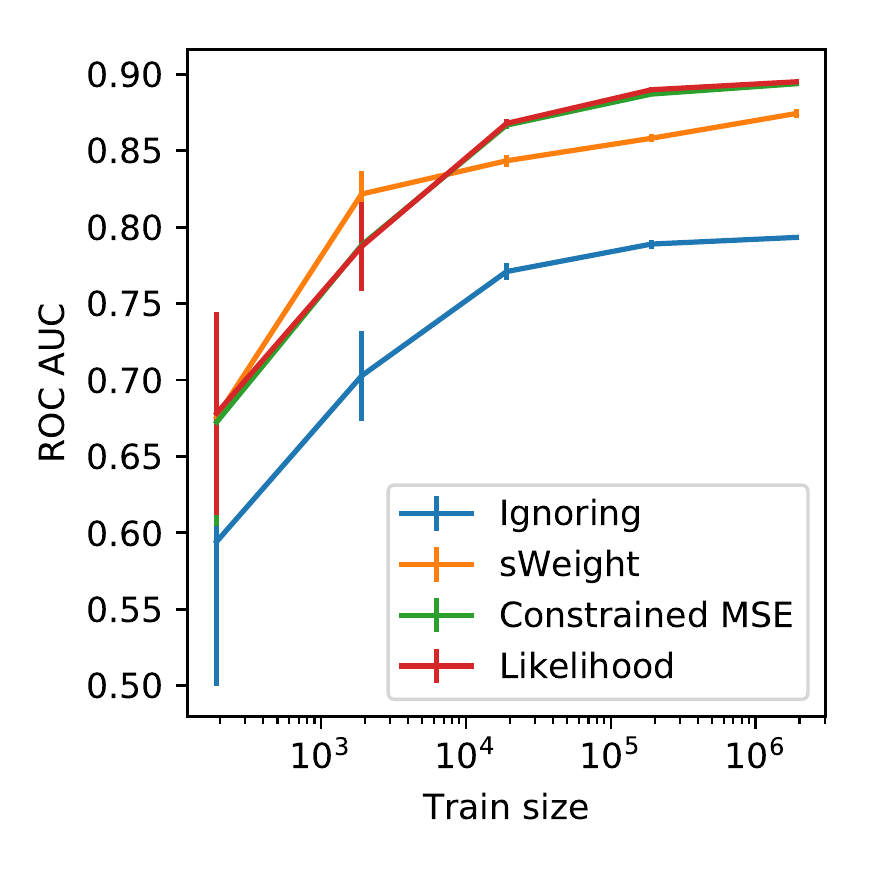}
        \caption{CatBoost}
        \label{fig:MuID-cb-our}
    \end{subfigure}
    \begin{subfigure}{0.5\textwidth}
        \includegraphics[width=\textwidth]{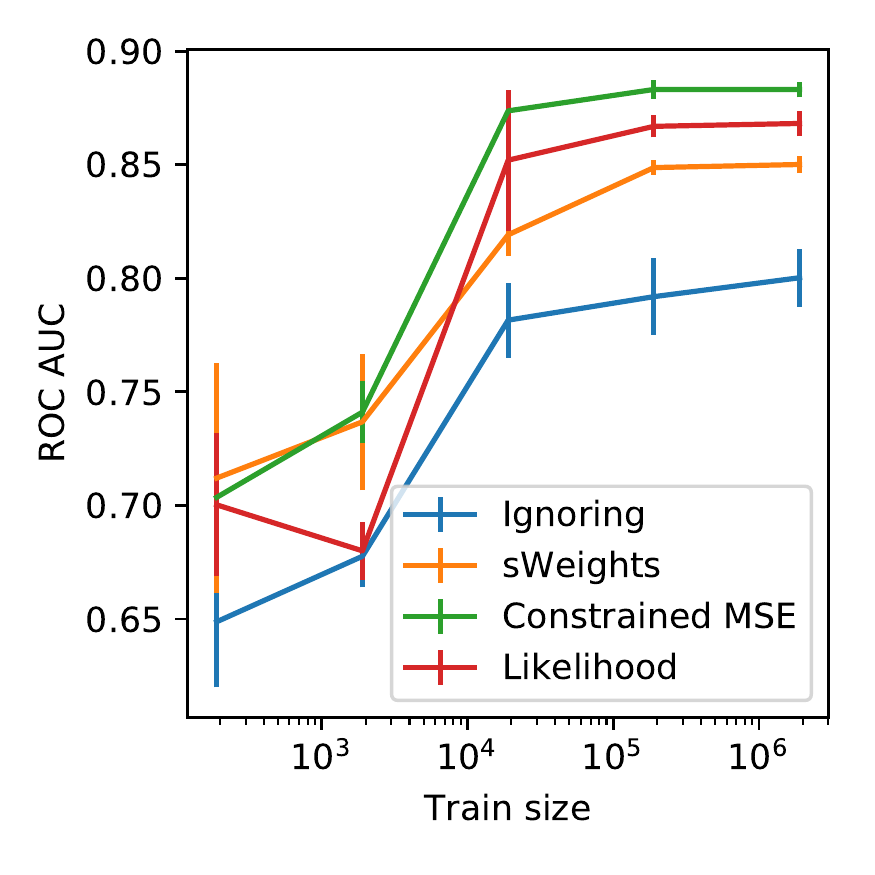}
        \caption{Deep Neural Network}
    \end{subfigure}
    \caption{Experimental evaluation of our loss functions on the MuID dataset with artificial sWeights. sWeights -- using sWeights directly as weights for logloss; Constrained MSE -- our regression (\ref{eq:constrainedMSE}) used to transform the sWeights, the result used as weights for classification; Likelihood -- our likelihood (\ref{eq:exactloss}) for the transformation; Ignoring weight -- training ignoring the sWeights.}
    \label{fig:results}
\end{figure}

As the result, we increase the impact of the sWeights on the classification problem to the point, where ignoring the sWeights leads to a significant drop in performance, while both our method and the baseline (using sWeights as example weights) are affected equally to permit for a fair comparison. The results are present in figure \ref{fig:results}. Surprisingly, for CatBoost with train size equal to $2\cdot10^3$, using sWeights as example weights has the best performance. In all other cases, our methods perform better than using sWeights as example weights during classifier training.

\section{Conclusion}
Training a machine learning algorithm on mixtures of signal and background events in the absence of exact labels is a challenging task. With discriminative variables available, the sPlot technique becomes a tempting solution. However, it might potentially lead to loss function with no lower bound, which in turn might result in catastrophic overfitting.

In this work, we consider the problem of separating two classes, each contaminated by background events. Building upon the prior work on signal-background separation, we introduce two loss functions that guarantee the stability of training by avoiding negative weights. Experimental evaluation shows improved classification quality compared to using sWeights directly as example weights, both for neural network and gradient boosting model

\section{Acknowledgments}
The research leading to these results has received funding from Russian Science Foundation under grant agreement n 17-72-20127.

We are grateful to the LHCb collaboration for opening their data; Artem Maevskiy for the weighted ROC AUC code.

\section{References}
\bibliographystyle{iopart-num}
\bibliography{bibliography}
\appendix

\section{Models parameters} \label{sec:model_parameters}
\begin{itemize}
    \item Fully-connected neural networks (NN): 4 layers, 512, 256, 128 neurons in the first three layers and either 1 or 2 neurons in the last one, leaky ReLu (0.05), optimized by adam \cite{kingma2014adam} algorithm with learning rate $2 \cdot 10^{-4}$, $\beta_1=0.9$, $\beta_2=0.999$, trained for 32 epochs; regressors and classifiers use the same architecture;
    \item CatBoost: 500 trees, leaf\_estimation\_method="Gradient", version 0.10.2 with our losses added and check for negative weights removed: \url{\detokenize{https://github.com/kazeevn/catboost/tree/constrained_regression}}
\end{itemize}
\end{document}